\documentclass[12pt]{article}
\usepackage{amssymb, amsmath, amsfonts}

\newtheorem{thm}{Theorem}[section]
\newtheorem{pro}[thm]{Proposition}

\newtheorem{rem}[thm]{Remark}
\newtheorem{defn}[thm]{Definition}
\newtheorem{ex}{Example}

\font\myfont=msbm10 scaled \magstep1

\def\ZZ{\hbox{\myfont\char'132}}
\def\NN{\hbox{\myfont\char'116}}

\begin{document}

\begin{center}
{\Large  Discretization of Liouville type nonautonomous equations preserving integrals }

{Ismagil Habibullin}\footnote{e-mail: habibullinismagil@gmail.com}\\

{Institute of Mathematics, Ufa Scientific Center, Russian Academy of Sciences,\\
Chernyshevskii Str., 112, Ufa, 450077, Russia\\
and\\
Bashkir State University, Z.Validi str. 32, Ufa, 450076, Russia}\\

\bigskip

{Natalya Zheltukhina}\footnote{e-mail: natalya@fen.bilkent.edu.tr}

{Department of Mathematics, Faculty of Science,
 \\Bilkent University, 06800, Ankara, Turkey \\}

\end{center}

\begin{abstract}
The problem of constructing semi-discrete integrable analogues of the Liouville type integrable PDE is discussed. We call the semi-discrete equation a discretization of the Liouville type PDE if these two equations have a common integral. For the Liouville type integrable equations from the well-known Goursat list for which the  integrals of  minimal order are of the order  less than or equal to two  we presented a list of corresponding semi-discrete versions. The list contains new examples of non-autonomous Darboux integrable chains.
\end{abstract}


\section{Introduction}

At the present time the problem of discretization of the integrable differential equations is
actively  studied. In the literature one can find various approaches and techniques used to
solve this problem including the  B\"{a}cklund transformation, the Hamiltonian structure,
symmetries, Lax pair, finite gap integration (see \cite{Ragnisco}-\cite{Adler}).
In our previous work \cite{Discretization} we considered the discretization of the Liouville type partial differential
equations preserving the structure of one of the integrals, and we constructed the semi-discrete
analogues for some equations found by E. Goursat \cite{Goursat}. However, semi-discrete analogues were
not found there for nonautonomous differential equations. Moreover, in \cite{Discretization} we did not
evaluate the continuum limit equations of the chains obtained by the discretization.

In the present paper we applied the discretization via integrals procedure to nonautonomous cases as well. We also discuss continuum limit equations for some particular
semi-discrete analogues obtained via the discretization. It is verified that discretization of a given Liouville
type PDE found by some formal manipulations after evaluation of the continuum limit for vanishing of the grid parameter $\varepsilon$
arrives at just the same PDE.

We  consider
semi-discrete chains of the form
\begin{equation}\label{dhyp}
\frac{d}{dx}t(n+1, x)=f(x,n,t(n, x),t(n+1, x),\frac{d}{dx}t(n, x))\, ,
\end{equation}
where unknown function $t=t(n,x)$ depends on discrete and continuous variables $n$ and $x$, respectively.
We use the following notations throughout the paper:
$$t_k=t(n+k,x),\quad  k\in \ZZ,\quad \qquad
t_{[m]}=\frac{d^m}{dx^m}t(n,x), \quad m\in \NN.
$$
 Denote by $D$ and $D_x$ the shift operator and the
operator of the total derivative with respect to $x$ correspondingly:
 $$Dh(n,x)=h(n+1,x), \qquad \qquad
D_xh(n,x)=\frac{d}{d x}h(n,x).$$
Let us recall the necessary definitions (see \cite{HZhP2008}, \cite{HZhP2009} for more details).
{\defn{Functions $I$ and $F$, depending on $x$, $n$,
$\{t_{[m]}\}_{m=1}^{\infty}$, $\{t_k\}_{k=-\infty}^{\infty}$, are called  respectively  $n$- and $x$-integrals
of (\ref{dhyp}), if  $DI=I$ and $D_xF=0$.}}

Any function depending on $n$ only, is an $x$-integral, and any
function, depending on $x$ only, is an $n$-integral.
Such integrals are called trivial integrals.
One can show  that any $n$-integral $I$ does
not depend on variables $t_{m}$ for $m\in \ZZ\backslash\{0\}$,
 and any $x$-integral $F$
does not depend on variables $t_{[m]}$ for $m\in \NN$.

{\defn{Chain (\ref{dhyp}) is called {\it Darboux integrable}
if it admits a nontrivial $n$-integral  and a nontrivial
$x$-integral.}}

Note that the order of the $n$-integral $I=I(n,x,t,t_x,...,D_x^mt)$ equals m. Starting with $I$ we can produce a new integral $H$ by setting
\begin{equation}\label{newintegral}
H=H(x,I,D_xI,...D_x^kI)
\end{equation}
Evidently its order is $k+m$. It can be proved that chain  (\ref{dhyp}) having a nontrivial integral admits a nontrivial integral of the minimal order which plays the key role: any $n$-integral $H$ can be represented in the form (\ref{newintegral}) through the minimal order $n$-integral $I$.

It can be verified that it is possible to find autonomous $x-$ and $y-$integrals of minimal order for any Liouville type equation of the form $u_{xy}=f(u,u_x,u_y)$, i.e. for an equation having no
explicit dependence on $x, y$. This fact is clearly illustrated by the list of equations found by E.Goursat in \cite{Goursat}.
In the recent paper  \cite{Yamilov}
the authors presented a class of discrete autonomous equations possessing both nontrivial integrals of
minimal orders depending on independent discrete variables. The existence of such examples, showing
that the class of discrete equations has more complicated structure, stimulated our interest to the
discretization problem.

Chain (\ref{dhyp}) is a semi-discrete analogue of the well-studied hyperbolic type equation
\begin{equation}\label{dhypcontinuous}
u_{xy}=g(x,y,u, u_x, u_y)\, .
\end{equation}

{\defn{Functions $W(x,y,u,u_x,u_{xx},...)$ and $\bar W(x,y,u,u_y,u_{yy},...)$, are called  respectively  $y$- and $x$-integrals
of (\ref{dhypcontinuous}), if  $D_yW=0$ and $D_x\bar W=0$.}}

{\defn{Equation    $u_{n+1,x}=f(x,n,u_n,u_{n+1},u_{n,x})$ is called a discretization of the equation (\ref{dhypcontinuous}) if these two equations have a common integral $W(x,y,u,u_x,u_{xx},...)\approx I(x,n,u_{n},u_{n,x},u_{n,xx}...)$.
Here the relation $W\approx I$ means that $I$ is obtained from $W$ by replacing $y\rightarrow n\varepsilon$, $u\rightarrow u_n$, $u_x\rightarrow u_{n,x}$, $u_{xx}\rightarrow u_{n,xx}$ and so on.}}

In \cite{Goursat} E.Goursat presented a list of Darboux integrable equations. We selected from the list only those equations for which the minimal order integrals have the orders no greater than 2. The trivial case when both
$x$-integral $W(x,y,u,u_y)$ and $y$-integral $\bar{W}(x,y,u,u_x)$ are
of order 1 is excluded:\\
(I) $u_{xy}=e^u$, $\bar{W}=u_{xx}-(1/2)u_x^2$,  $W=u_{yy}-(1/2)u_y^2$;\\
(II) $u_{xy}=e^u u_y$, $\bar{W}=u_x-e^u$, $W=\frac{u_{yy}}{u_y}-u_y$;\\
(III) $u_{xy}=e^u\sqrt{u_y^2-4}$, $\bar{W}=u_{xx}-(1/2)u_x^2-(1/2)e^{2u}$, $W=\frac{u_{yy}-u_y^2+4}{\sqrt{u_y^2-4}}$;\\
(IV) $u_{xy}=u_xu_y\left(\frac{1}{u-x}+\frac{1}{u-y}\right)$, $\bar{W}=\frac{u_{xx}}{u_x}-\frac{2u_x}{u-x}+\frac{1}{u-x}$, $W=\frac{u_{yy}}{u_y}-\frac{2u_y}{u-y}+\frac{1}{u-y}$;\\
(V) $u_{xy}=\psi(u)\beta(u_x)\bar{\beta}(u_y)$, $(ln\psi)''=\psi^2$, $\beta\beta'=-u_x$,
$\bar{\beta}\bar{\beta}'=-u_y$, \\
 $\qquad \qquad \bar{W}=\frac{u_{xx}}{\beta(u_x)}-\frac{\psi'(u)}{\psi(u)}\beta(u_x)$, $W
=\frac{u_{yy}}{\bar{\beta}(u_y)}-\frac{\psi'(u)}{\psi(u)}\bar{\beta}(u_y)$;\\
(VI) $u_{xy}=\frac{\beta(u_x)\bar{\beta}(u_y)}{u}$, $\beta\beta'+c\beta=-u_x$, $\bar{\beta}\bar{\beta}'+c\bar{\beta}=-u_y$,\\
$ \qquad \qquad \bar{W}=\frac{u_{xx}}{\beta}-\frac{\beta}{u}$, $W=\frac{u_{yy}}{\bar{\beta}}-\frac{\bar{\beta}}{u}$;\\
(VII) $u_{xy}=-2\frac{\sqrt{u_xu_y}}{x+y}$, $\bar{W}=\frac{u_{xx}}{\sqrt{u_x}}+2\frac{\sqrt{u_x}}{x+y}$,
$W=\frac{u_{yy}}{\sqrt{u_y}}+2\frac{\sqrt{u_y}}{x+y}$;\\
(VIII) $u_{xy}=\frac{1}{(x+y)\beta(u_x)\bar{\beta}(u_y)}$, $\beta'=\beta^3+\beta^2$,
$\bar{\beta}'=\bar{\beta}^3+\bar{\beta}^2$,\\
$\qquad \qquad \bar{W}=u_{xx}\beta(u_x)-\frac{1}{(x+y)\beta(u_x)}$,
$W=u_{yy}\bar{\beta}(u_y)-\frac{1}{(x+y)\bar{\beta}(u_y)}$.

Throughout the paper we shortly call the list of eight equations above as the Goursat list.
 Note that the work \cite{Goursat} contains also equations for which the minimal order integrals are of the order higher than two.

According to Definition 1.4 in order to discretize a Darboux integrable equation of the form (\ref{dhypcontinuous}) we have to solve a kind of the inverse problem: search the equation of the form (\ref{dhyp}) possessing the given integral.

In \cite{Discretization} we made a discretization of equations (\ref{dhypcontinuous}) preserving the structure of
$y$-integrals in each of eight equations from the list (I)-(VIII).
The discretization in  \cite{Discretization}, where it is supposed that $n$-integrals are functions not depending on $n$,
did not provide semi-discrete equations for each function $\beta(t_x)$ in three cases, namely cases V, VI and VIII.
Also, in cases IV and VII, where $y$-integrals depend on $x$ and $y$, the obtained in  \cite{Discretization}
semi-discrete chains did not have the corresponding continuous limit equations.

\section{Statements of the results}

In the present paper we allow $n$-integral and function $f$ explicitly depend on $x$ and $n$, and with this
modification  in the discretization algorithm we again study all cases I - VIII.
In cases V, VI  and VIII the  $n$-integrals depend on functions $\beta$ that are solutions of some
differential equations.
Below we give semi-discrete versions of these  equations in the Goursat list.

{\thm\label{newDiscretizationcase5} (Case V) { Semi-discrete chain $t_{1x}=f(x,t,t_1,t_x)$ possessing a minimal order $n$-integral $I=\frac{t_{xx}}{\beta(t_x,n)}+\frac{\psi'(t,n)}{\psi(t,n)}\beta(t_x,n)$, where
$(\ln \psi)''=\psi^2$ and $\beta'(t_x,n)\beta(t_x,n)=-t_x$ is
$$t_{1x}=\lambda(t,t_1,n)t_x+\mu(t,t_1, n)\beta(t_x,n)\, $$ with $\lambda$ and $\mu$ satisfying the equations
$$\lambda^2+\mu^2=\nu(n)\, ,\quad
\lambda_{t_1}-\frac{\psi'(t_1, n+1)}{\psi(t_1,n+1)}\lambda +\frac{\psi'(t,n)}{\psi(t,n)}=0\, ,\quad  \lambda_{t}-\frac{\psi'(t, n)}{\psi(t,n)}\lambda +\nu\frac{\psi'(t_1,n+1)}{\psi(t_1,n+1)}=0\, ,$$
 where  $\nu(n)$ is some constant depending on $n$ only.\\
 This semi-discrete chain has $x$-integral  $F=\psi(t_1,n+1) E(t,t_1,t_2)$, where $E_t=\frac{1}{\mu(t,t_1, n)}$, $E_{t_2}=\frac{1}{\nu \mu(t_1,t_2, n+1)}$ and
$E_{t_1}=-\frac{\lambda(t_1,t_2, n+1)}{\nu \mu(t_1, t_2, n+1)}-\frac{ \lambda(t,t_1, n)}{\nu \mu(t, t_1, n)}-\frac{\psi'(t_1, n+1)}{\psi(t_1,n+1)}E$.

} }

\bigskip

Note that the overdetermined systems of the differential equations for defining $\lambda$ and, respectively $E$, are compatible (see  section 3 below).

\bigskip

{\thm\label{newDiscretizationcase6}(Case VI){ Semi-discrete chain $t_{1x}=f(n,t,t_1,t_x)$ possessing a minimal order $n$-integral
$I=\frac{t_{xx}}{\beta(t_x,n)}-\frac{\beta(t_x,n)}{t}$, where
$\beta'(t_x,n)\beta(t_x,n)+C\beta(t_x,n)=-t_x$ is
$$t_{1x}=\lambda(t,t_1,n)t_x+\mu(t,t_1, n)\beta(t_x,n)\, $$ with
\begin{equation}\label{lambda}\left\{ \begin{array}{l}
\lambda_t=\frac{\mu^2+\lambda^2-C\lambda \mu}{t_1}-\frac{\lambda }{t}\, ,\\
\lambda_{t_1}=\frac{C\mu -\lambda}{t_1}+\frac{1}{t}\, ,
\end{array} \right.
\end{equation}
where
$$\left\{
\begin{array}{ll}
(B\lambda-\mu)^{-B^2}(\lambda-B\mu)=\nu(n), \quad  B=\frac{C-\sqrt{C^2-4}}{2},&
\qquad {\mbox{if}} \qquad C^2>4,\\
\ln(\lambda^2-C\lambda \mu +\mu^2)-\frac{2C}{\sqrt{4-C^2}}\arctan \frac{2\lambda
-C\mu}{\mu\sqrt{4-C^2}}=\nu(n), & \qquad {\mbox{if}} \qquad C^2<4,\\
\ln(\lambda -\mu)+\frac{\mu}{\lambda -\mu}=\nu(n), &\qquad {\mbox{if}} \qquad C=2,\\
\ln(\lambda +\mu)-\frac{\mu}{\lambda +\mu}=\nu(n), &\qquad {\mbox{if}} \qquad C=-2,
\end{array}\right.
$$
and $\nu(n)$ is some constant depending on $n$ only.\\
This semi-discrete chain has $x$-integral  $F=\frac{1}{t_1} E(t,t_1,t_2)$, where $E_{t_2}=\frac{1}{\mu(t_1,t_2, n+1)}$ ,\\
$E_t=\frac{\mu^2(t,t_1,n)+\lambda^2(t,t_1,n)-C\lambda(t,t_1,n)
\mu(t,t_1,n)}{\mu(t,t_1, n)}$ and
$E_{t_1}=-\frac{\lambda(t_1,t_2, n+1)}{\mu(t_1, t_2, n+1)}-\frac{\lambda(t,t_1,
n)}{\mu(t, t_1, n)}+C+\frac{1}{t_1}E$.

} }

{\thm\label{newDiscretizationcase8}(Case VIII){ Semi-discrete chain $t_{1x}=f(x,n,t,t_1,t_x)$ possessing a minimal order $n$-integral $I={\beta(t_x,n)}t_{xx}-\frac{1}{(x+\alpha(n))\beta(t_x,n)}$, where
$\beta'(t_x,n)=\beta^3(t_x,n)+\beta^2(t_x,n)$ and  $\alpha(n)$ is some constant depending on $n$ only, is
$$t_{1x}=\frac{1-K}{\beta(t_x,n)}+t_x+(-K+\ln K)  \, $$ with function $K(x,n,t,t_1)$ satisfying the following system of equations
\begin{equation}\label{functionK}\left\{ \begin{array}{l}
K_t+K_{t_1}=0\, ,\\
K_{t_1}=\frac{K}{K-1}\left\{\frac{K}{x+\alpha(n+1)}-\frac{1}{x+\alpha(n)}\right\}\, ,\\
K_x=\frac{K}{K-1}\left\{\frac{K}{x+\alpha(n+1)}-\frac{1}{x+\alpha(n)}\right\}(K-\ln K)-\frac{(K-1)K}{x+\alpha(n+1)}\, .
\end{array} \right.
\end{equation}
This semi-discrete chain has $x$-integral  $F=\frac{1}{x+\alpha(n+1)} E(x,t,t_1,t_2)$, where
$E_x=\frac{K(1-\ln K)}{1-K}-\frac{1-\ln K_1}{1-K_1}+\frac{1}{x+\alpha(n+1)}E$, $E_{\tau_1}=\frac{K}{1-K}$, $E_{\tau_2}=-\frac{1}{1-K_1}$ with
$\tau_1=t_1-t$ and $\tau_2=t_2-t_1$.
} }

\bigskip

\noindent Let us now present  one particular case described in Theorem \ref{newDiscretizationcase5} corresponding to $\beta(t_x)=\sqrt{1-t_x^2}$ and $\psi(t)=-\frac{1}{t}$.

{\ex{  Semi-discrete chain
\begin{equation}\label{example}
t_{1x}=\frac{t_1^2+\nu(n) t^2}{2tt_1}t_x+i\, \frac{\nu(n)t^2-t_1^2}{2tt_1}\sqrt{1-t_x^2}
\end{equation}
 has $n$-integral $I=\frac{t_{xx}}{\sqrt{1-t_x^2}}-\frac{\sqrt{1-t_x^2}}{t}$ and $x$-integral
$F=\frac{\nu(n)t_1^2-t_2^2}{\nu(n)t^2-t_1^2}$ for any constant  $\nu(n)$  depending on $n$ only. If in (\ref{example}) one substitutes $u$ and $u+\varepsilon e^{\gamma(u_y)}$ with $\gamma'=1/\beta$, instead of $t$ and $t_1$ correspondingly, and let $\varepsilon$ approach $0$, continuous Liouville equation analogue $u_{xy}=\frac{\beta(u_x)\beta(u_y)}{u}$ would be obtained.
 }}

In cases IV and VII the  $y$-integrals depend on the variables $x$ and $y$. We consider
these special nonautonomous  cases,  allowing explicit $n$-dependence of $n$-integral and of the function $f$, and  obtain some new semi-discrete chains.

\begin{thm}\label{newDiscretization}(Cases IV and VII)
\noindent (a) Semi-discrete equation (\ref{dhyp}) possessing an $n$-integral \\$\displaystyle{I=\frac{t_{xx}}{t_x}-\frac{2t_x}{t-x}+\frac{1}{t-x}}$
is
\begin{equation}\label{Case4new}
t_{1x}=\frac{(1+t_1M(n))(t_1-x)}{(1+tM(n))(t-x)}t_x\,
\end{equation}
where $M(n)$ is an arbitrary function of $n$. Function
$\displaystyle{F=\frac{(1+t_2M(n+1))(t_1-t)}{(1+tM(n))(t_1-t_2)}}$ is an $x$-integral of (\ref{Case4new}).

\noindent(b) Semi-discrete equation (\ref{dhyp}) possessing an $n$-integral
 $\displaystyle{I=\frac{t_{xx}}{\sqrt{t_x}}+
\frac{2\sqrt{t_x}}{x+\varepsilon n}}$ is
\begin{equation}\label{Case7new}
t_{1x}=(\sqrt{t_x}+\alpha)^2\, ,\, \,\,  \alpha=
\sqrt{\frac{\varepsilon(t_1-t)}{(x+\varepsilon n)(x+\varepsilon (n+1))}}\, .
\end{equation}
Function $\displaystyle{F=(x+\varepsilon n)
\alpha-(x+\varepsilon (n+2))D\alpha }$
is an $x$-integral of (\ref{Case7new}).
\end{thm}

\begin{thm} \label{ContinuumLimits}(Cases I-IV and VII)
Below we display continuum limit equations and $x$-integrals for semi-discrete equations obtained by discretization of the continuous equations from the Goursat list.
$$
\begin{array}{|l|l|l|}
\hline
Semi-discrete\,\, equation & Continuum\,\, limit\,\, equations& \\
and\, \,its\,\, x-integral\,\,F&and \,\,x-integrals \,\,\tilde{F}&\\
\hline
t_{1x}=t_x+Ce^{(1/2)(t+t_1)}, \, C=\varepsilon & u_{xy}=e^u&A\\
F=e^{(t_1-t)/2}+e^{(t_1-t_2)/2}&\lim\limits_{\varepsilon\to 0}2\varepsilon^{-2}(2-F)=u_{yy}-(1/2)u_y^2=\tilde{F}&\\
\hline
t_{1x}=t_x-e^t+e^{t_1}& u_{xy}=e^{u}u_y&B\\
F=(e^{t}-e^{t_2})(e^{t_1}-e^{t_3})(e^{t}-e^{t_3})^{-1}(e^{t_1}-e^{t_2})^{-1}&\lim\limits_{\varepsilon\to 0}
\frac{12}{\varepsilon^{2}}(1-F)=-2\tilde{F}_y+\tilde{F}^2, \,&\\
& \tilde{F}=\frac{u_{yy}}{u_y}-u_y&\\
\hline
t_{1x}=K(t,t_1)t_x, \, K=1+\varepsilon e^{t_{1}}&u_{xy}=e^{u}u_x&C\\
F=e^{t-t_1}+\varepsilon e^t&\lim\limits_{\varepsilon\to 0}\varepsilon^{-1}(1-F)=u_y-e^u=\tilde{F}&\\
\hline
t_{1x}=t_x+\sqrt{e^{2t}+Re^{t+t_1}+e^{2t_1}}, R=-2-4\varepsilon^2&u_{xy}=e^u\sqrt{u_y^2-4}&D\\
F= arcsinh(ae^{t_1-t_2}+b)+arcsinh(ae^{t_1-t}+b)& \lim\limits_{\varepsilon\to 0}\varepsilon^{-1}(-F+4\ln 2)=\frac{u_{yy}-2u_y^2+4}{\sqrt{u_y^2-4}}=\tilde{F}&\\
a= (-4\varepsilon^4-4\varepsilon^2)^{-1/2},\, b=-(1+2\varepsilon^2)a &&\\
\hline
t_{1x}=\sqrt{R^2e^{2(t+t_1)}+2Re^{t+t_1}}\sqrt{t_x^2-4}+&u_{xy}=e^u\sqrt{u_x^2-4}&E\\
\hskip1cm (1+Re^{t+t_1})t_x\,,R=2^{-1}\varepsilon^2 &&\\
F=\sqrt{Re^{2t_1}+2e^{t_1-t}}+\sqrt{Re^{2t_1}+2e^{t_1-t_2}}&\lim\limits_{\varepsilon\to 0}\frac{1}{\varepsilon^{2}}
(4-\sqrt{2}F)=u_{yy}-\frac{1}{2}(u_y^2+e^{2u})&\\
\hline
t_{1x}=\frac{(1+t_1M(n))(t_1-x)}{(1+tM(n))(t-x)}t_x,\, M=-\frac{1}{\varepsilon n} & u_{xy}=u_xu_y\left(\frac{1}{u-x}+\frac{1}{u-y}\right)&F\\
F=\frac{(1+t_2M(n+1))(t_1-t)}{(1+tM(n))(t_1-t_2)}&\lim\limits_{\varepsilon\to 0}\frac{1}{\varepsilon}
((1+n^{-1})F+1)=\frac{1-2u_y}{u-y}+\frac{u_{yy}}{u_y}&\\
\hline
t_{1x}=(\sqrt{t_x}+\alpha)^2, \,
\alpha =\sqrt{\frac{\varepsilon(t_1-t)}{(x+\varepsilon n)(x+\varepsilon (n+1))}}&
u_{xy}=2\frac{\sqrt{u_xu_y}}{x+y}&G
\\
F=(x+\varepsilon n)\alpha-(x+\varepsilon (n+2))D\alpha &
\lim\limits_{\varepsilon\to 0}\frac{-F}{\varepsilon^2}=\frac{\sqrt{u_y}}{x+y}+\frac{1}{2}\frac{u_{yy}}{\sqrt{u_y}}&\\
\hline
\end{array}
$$
\end{thm}

\bigskip

In the present paper we concentrate mainly on the ``discretization" i.e. on the evaluation of the discrete versions preserving the
structure of the integrals. The inverse operation is also meaningful. According to Definition 1.4 we can look for PDE of the form (\ref{dhypcontinuous}) starting with the known integral of a Darboux
integrable chain (\ref{dhyp}). Another way to find the continuous counterpart is connected with the  evaluating the continuum limit. Remark that these two methods give one and the same answer. Let us give an illustrative example.

{\rem{ Let us find  all equations $t_{xy}=f(x,y, t,t_x,t_y)$ possessing a $y$-integral
$I=t_{xx}-(1/2)t_x^2$, that is, we are looking for a continuous analogue of semi-discrete chain
$t_{1x}=t_x+Ce^{(1/2)(t+t_1)}$ (case ($A$)) preserving the structure of its $n$-integral.
Equality $D_y I=0$ becomes $t_{xxy}-t_x t_{xy}=0$. From the equation searched $t_{xy}=f(x,y, t,t_x,t_y)$
we obtain $t_{xxy}=f_x+f_t t_x+f_{t_x}t_{xx}+f_{t_y}f$. Therefore,
\begin{equation}\label{ex1}
f_x+f_t t_x+f_{t_x}t_{xx}+f_{t_y}f-t_x f=0\, .
\end{equation}
Evidently, the coefficient before $t_{xx}$ in (\ref{ex1}) vanishes, that is $f_{t_x}=0$.
Now collection of the coefficients before $t_x$ in (\ref{ex1}) gives $f_{t}-f=0$, or $f=A(x,y,t_y)e^t$.
 We substitute the expression  $f=A(x,y,t_y)e^t$ into (\ref{ex1}) and get $A_xe^t+A_{t_y}e^{2t}=0$ which
immediately implies $A_x=A_{t_y}=0$. Therefore, the equation searched is of the form $t_{xy}=A(y)e^t$ which
coincides with the Liouville equation up to a point transformation $y\to \tilde{y}=\int_0^y A(\theta)\, d\theta$.
 It is remarkable that usual continuum limit with small $\varepsilon =C>0$ approaching zero gives the same answer:
equation $(t_{1x}-t_x)/\varepsilon=e^{(1/2)(t+t_1)}$ becomes the Liouville equation.}}

Remark convinces that the problem of evaluating the PDE by its known integral is trivially solved. For the semi-discrete chain it is not the case.
The matter is that in this case instead of the differential relation $D_yW=0$ we have a functional equation $DI=I$.

It is widely known that integrable discretization is closely connected with the B\"acklund transformation.
We discuss this connection in section \ref{BacklandDiscretization}. It is shown  that some of the discrete models coincide with the B\"{a}cklund transformation for the continuous counterparts, while the others do not.

We prove Theorems \ref {newDiscretizationcase5} -
 \ref{newDiscretization} in sections \ref{THV} -
  \ref{THIV,VII}, and present the proof of
Theorem \ref{ContinuumLimits}
  in two special cases $F$ and $G$ in section \ref{C}. Other cases from
Theorem \ref{ContinuumLimits}
 can be proved in a similar way.

\section{Proof of Theorem \ref{newDiscretizationcase5}\label{THV}}

\noindent {\bf{\underline{Discretization}}}: Let us find all chains $t_{1x}=f(x,n,t,t_1,t_x)$ with $n$-integral $I=\frac{t_{xx}}{\beta(t_x,n)}+\frac{\psi'(t,n)}{\psi(t,n)}\beta(t_x,n)$, where
\begin{equation}\label{initial}(\ln \psi)''=\psi^2, \qquad  \beta'(t_x,n)=-\frac{t_x}{\beta(t_x,n)}, \qquad \beta'(f,n+1)=-\frac{f}{\beta(f,n+1)}\, .
\end{equation}
$DI=I$ implies
\begin{equation}\label{1}
\frac{f_x+f_t t_x+f_{t_1}f+f_{t_x}t_{xx}}{\beta(f,n+1)}+\frac{\psi'(t_1,n+1)}{\psi(t_1,n+1)}\beta(f,n+1)=\frac{t_{xx}}{\beta(t_x,n)}+\frac{\psi'(t,n)}{\psi(t,n)}\beta(t_x,n)\, .
\end{equation}
We compare the coefficients before $t_{xx}$ and get
\begin{equation}\label{derivative}
\frac{f_{t_x}}{\beta(f,n+1)}=\frac{1}{\beta(t_x,n)}\, ,
\end{equation} or
$$
\gamma(f,n+1)=\gamma (t_x,n)+A(x,n,t,t_1), \qquad  {\mbox{where }}\qquad \gamma'(t_x,n)=\frac{1}{\beta(t_x,n)}\, .
$$
We have, $\gamma'(f,n+1)f_{t_1}=A_{t_1}$, or $f_{t_1}=A_{t_1}\beta(f,n+1)$. Similarly,
 $f_{t}=A_{t}\beta(f,n+1)$ and $f_{x}=A_{x}\beta(f,n+1)$. Substitute these expressions for $f_{x}$, $f_{t}$ and $f_{t_1}$ into
(\ref{1}) and get
\begin{equation}\label{2}
A_x+t_xA_t+A_{t_1}f+r_1\beta(f,n+1)=r\beta(t_x,n)\, ,
\end{equation}
where
\begin{equation}\label{r}
 r=\frac{\psi'(t,n)}{\psi(t,n)}, \qquad  r_1=\frac{\psi'(t_1,n+1)}{\psi(t_1,n+1)}\, .\end{equation}
Differentiate with respect to $t_x$ equality (\ref{2}), use (\ref{derivative}) and (\ref{initial}), and get
\begin{equation}\label{3}
A_t\beta(t_x,n)+A_{t_1}\beta(f,n+1)-r_1f=-rt_x\, .
\end{equation}
Differentiate with respect to $t_x$ equality (\ref{3}), use (\ref{derivative}) and  (\ref{initial}), and obtain
\begin{equation}\label{4}
t_xA_t +A_{t_1}f+r_1\beta(f,n+1)=r\beta(t_x,n)\, .
\end{equation}
One can see from (\ref{2}) and (\ref{4}) that $A_x=0$.   We express $\beta(f,n+1)$ from (\ref{4}), substitute it into (\ref{3}) and get
\begin{equation}\label{f}
f=\lambda t_x+\mu\beta(t_x, n)\, ,
\end{equation}where
\begin{equation}\label{coefficients}
\lambda=\frac{rr_1-A_tA_{t_1}}{r_1^2+A_{t_1}^2}, \qquad \mu=\frac{r_1A_t+rA_{t_1}}{r_1^2+A_{t_1}^2\, .}
\end{equation}
Note that $f_{t_x}=\lambda +\mu\beta'(t_x,n)=\lambda-\mu\frac{t_x}{\beta(t_x,n)}$ by (\ref{f}) and (\ref{initial}). On the other hand, $f_{t_x}=\frac{\beta(f,n+1)}{\beta(t_x,n)}$, by (\ref{derivative}). Hence,
\begin{equation}\label{beta}
\beta(f,n+1)=-\mu t_x+\lambda \beta(t_x,n)\, .
\end{equation}
It follows from (\ref{initial}) that
\begin{equation}\label{square}
\beta^2(t_x,n)=-t_x^2+C(n), \qquad \beta^2(f,n+1)=-f^2+C(n+1),
\end{equation}
where $C(n)$ and $C(n+1)$ are some constants. Since
$$
f^2=\lambda^2t_x^2+2\lambda\mu t_x\beta(t_x,n)+\mu^2\beta^2(t_x,n)\, ,
$$
$$
\beta^2(f,n+1)=\lambda^2\beta^2(t_x,n)-2\lambda \mu t_x\beta(t_x,n)+\mu^2t_x^2\, ,
$$
then $$
f^2+\beta^2(f, n+1)=(\lambda^2+\mu^2)(t_x^2+\beta^2(t_x, n))\, ,
$$
and, therefore, due to (\ref{square}),
\begin{equation}\label{nu}
\lambda^2+\mu^2=\nu\, ,
\end{equation}
where $\nu=C(n+1)/C(n)$ is some constant depending on $n$ only.

Let us show that \begin{equation}\label{A}
r^2+A_t^2=\nu(r_1^2+A_{t_1}^2)\, .
\end{equation}
Indeed,
$$\nu=\lambda^2+\mu^2=\frac{r^2r_1^2+A_t^2A_{t_1}^2+r_1^2A_t^2+r^2A_{t_1}^2}{(r_1^2+A_{t_1}^2)^2}
$$
can be rewritten as
$$
\nu(A_{t_1}^2)^2+(2\nu r_1^2-r^2-A_t^2)A_{t_1}^2+(\nu r_1^4-r^2r_1^2-r_1^2A_t^2)=0\, ,
$$
that implies
$$A_{t_1}^2=\frac{-(2\nu r_1^2-r^2-A_t^2)+r^2+A_t^2}{2\nu}\, ,
$$ that is equivalent to (\ref{A}).

We substitute expressions $f=\lambda t_x+\mu\beta(t_x,n)$ and $\beta(f, n+1)=-\mu t_x+\lambda \beta(t_x,n)$ into (\ref{1}) and get
$$
\lambda_t t_x^2+\mu_t\beta(t_x,n) t_x +(\lambda_{t_1}t_x+\mu_{t_1}\beta(t_x,n))(\lambda t_x+\mu\beta(t_x,n))
$$
$$
=(r\beta(t_x,n)-r_1\lambda\beta(t_x,n)+r_1\mu t_x)(\lambda \beta(t_x,n)-\mu t_x)\, .
$$
In the last equality we first replace $\beta^2(t_x, n)$ by $-t_x^2+C(n)$  due to (\ref{square}), and then we compare the coefficients before linearly independent functions
$t_x^0$, $t_x^2$ and $t_x\beta(t_x, n)$. We obtain,
\begin{equation}\label{1.1}
\mu_{t_1} \mu =\lambda r-\lambda^2r_1\, ,
\end{equation}
\begin{equation}\label{1.2}
\lambda_t+\lambda_{t_1}\lambda -\mu_{t_1}\mu=-\lambda r+\lambda^2r_1-\mu^2r_1\, ,
\end{equation}
and
\begin{equation}\label{1.3}
\mu_t+\lambda_{t_1}\mu +\mu_{t_1}\lambda =-\mu r +2 \lambda \mu r_1\, .
\end{equation}
Since $\lambda^2+\mu^2=\nu$, then $\mu_{t_1}\mu +\lambda_{t_1}\lambda=0$, and equation (\ref{1.1}) becomes
\begin{equation}\label{2.1}
\lambda_{t_1}-r_1\lambda +r=0\, .
\end{equation}
We subtract  (\ref{1.1}) from (\ref{1.2}), use (\ref{2.1}) and (\ref{nu}), and get
\begin{equation}\label{2.2}
\lambda_t-r\lambda +\nu r_1=0\, .
\end{equation}
One can check that equations (\ref{1.1})-(\ref{1.3}) are satisfied  if and only if equations (\ref{2.1}) and (\ref{2.2}) hold.
Note that equations (\ref{2.1}) and (\ref{2.2}) are compatible, since $\lambda_{tt_1}=\lambda_{t_1t}$ is equivalent to $\nu(r_1^2-\psi_1^2)=r^2-\psi^2$. The last one holds because
$(r^2-\psi^2)'=2rr'-2\psi\psi'=2\frac{\psi'}{\psi}\psi^2-2\psi\psi'=0$ as $r'=\psi^2$ by (\ref{initial}).

One can solve the system of equations  (\ref{2.1}) and (\ref{2.2}) and get that
$$
\lambda=\nu B(t)B(t_1) \psi(t)\psi(t_1)(\psi^2(t)-r^2(t)+C_1(n))-r(t)B(t_1)\psi(t_1)-\nu r(t_1)B(t)\psi(t)+C_2(n)\psi(t)\psi(t_1),
$$
where $B'=1/\psi$.

Note that equation (\ref{f})  can be written also as
\begin{equation}\label{ff}
\gamma(f,n+1)=\gamma (t_x,n)+A(t,t_1, n),
\end{equation}
where, due to (\ref{coefficients}) and (\ref{A}), we have
$$
\gamma'(t_x,n)=\frac{1}{\beta(t_x,n)}, \qquad A_{t_1}^2=\frac{(r-\lambda r_1)^2}{\nu-\lambda^2}, \qquad A_t^2=\frac{(\nu r_1-\lambda r)^2}{\nu-\lambda^2}\, ,$$
and $\lambda$ satisfies  (\ref{2.1}) and (\ref{2.2}).

\noindent{\bf{\underline{Finding $x$-integral}}}: Now we are looking for an $x$-integral $F(t,t_1,t_2)$ of equation (\ref{f})  satisfying (\ref{nu}), (\ref{2.1}) and  (\ref{2.2}). Equality $D_xF=0$ implies
$$
F_t t_x+F_{t_1}(\lambda t_x+\mu \beta(t_x,n))+F_{t_2}((\lambda_1\lambda -\mu_1\mu)t_x+(\lambda_1\mu+\mu_1\lambda)\beta(t_x,n))=0\,.
$$
By comparing the coefficients before $t_x$ and $\beta(t_x,n)$ in the last equation  we get the system of two equations
$$\left\{
\begin{array}{l}
F_t+\lambda F_{t_1}+(\lambda_1\lambda -\mu_1\mu)F_{t_2}=0\, ,\\
\, \qquad \mu F_{t_1}+(\lambda_1\mu+\mu_1\lambda)F_{t_2}=0\, ,
\end{array}\right.
$$
that can be rewritten as
\begin{equation}\label{system}\left\{
\begin{array}{l}
F_t\quad-\quad \nu\frac{\mu_1}{\mu}F_{t_2}=0\, ,\\
F_{t_1}+(\lambda_1+\lambda\frac{\mu_1}{\mu})F_{t_2}=0\, .
\end{array}\right.
\end{equation}
Note that $\left[ \frac{\partial }{\partial t}-\nu\frac{\mu_1}{\mu}\frac{\partial }{\partial t_2},\frac{\partial }{\partial t_1}+(\lambda_1+\lambda\frac{\mu_1}{\mu})\frac{\partial }{\partial t_2}\right]=
\left(\mu_1\left(\frac{\lambda}{\mu}\right)_t-\nu\mu\left(\frac{\lambda_1}{\mu_1}\right)_{t_2}+\nu\left(\frac{\mu_1}{\mu}\right)_{t_1}\right)\frac{\partial}{\partial t_2}=0$ due to (\ref{nu}), (\ref{2.1}) and  (\ref{2.2}).
Define function $E(t,t_1,t_2)$ to be such that
\begin{equation}\label{E}
E_t=\frac{1}{\mu}\, ,\qquad E_{t_2}=\frac{1}{\nu\mu_1}\, , \qquad E_{t_1}=-\frac{\lambda_1}{\nu\mu_1}-\frac{\lambda}{\nu\mu}-r_1E\, .
\end{equation}
 Such function exists since $E_{tt_2}=0=E_{t_2 t}$ and
$E_{tt_1}=E_{t_1 t}$, $E_{t_1 t_2}=E_{t_2 t_1}$ by (\ref{nu}), (\ref{2.1}) and  (\ref{2.2}). In new variables $\tilde{t}=E(t,t_1,t_2)$, $\tilde{t_1}=t_1$, $\tilde{t_2}=t_2$ system (\ref{system}) becomes
$$\left\{
\begin{array}{l}
F_{\tilde{t_2}}=0\, ,\\
 F_{\tilde{t_1}}-\frac{\psi'(\tilde{t_1})}{\psi(\tilde{t_1})}\, \tilde{t}F_{\tilde{t}}=0\, .
\end{array}\right. ,
$$
that implies that $x$-integral can be taken as $F(t,t_1,t_2)=\psi(t_1)E(t,t_1,t_2)$, where $E$ satisfies (\ref{E}).

\section{Proof of Theorem \ref{newDiscretizationcase6}\label{THVI}}

\noindent {\bf{\underline{Discretization}}}: Consider chains $t_{1x}=f(n, t, t_1, t_x)$ with $n$-integral
$I=\frac{t_{xx}}{\beta(t_x,n)}-\frac{\beta(t_x,n)}{t}$, where
\begin{equation}
\label{beta6}
\beta(t_x,n)\beta'(t_x,n)+C\beta(t_x,n)=-t_x.
\end{equation}
Equality $D I=I$ implies
\begin{equation}\label{2-6}
\frac{f_t t_x+f_{t_1}f+f_{t_x}t_{xx}}{\beta(f, n+1)}-\frac{\beta(f, n+1)}{t_1}=\frac{t_{xx}}{\beta(t_x,n)}-\frac{\beta(t_x,n)}{t}\, .
\end{equation}
By comparing the coefficients before $t_{xx}$ in (\ref{2-6}) we get
\begin{equation}\label{3-6}
\frac{f_{t_x}}{\beta(f,n+1)}=\frac{1}{\beta(t_x,n)}\, , \quad {\mbox{or}} \quad \gamma(f, n+1)-\gamma(t_x,n)=A(t,t_1, n) \quad {\mbox{with }} \gamma'=1/\beta\, .
\end{equation}
It follows from (\ref{3-6}) that $f_t=\beta(f,n+1)A_t$ and $f_{t_1}=\beta(f,n+1)A_{t_1}$. We substitute these expressions for $f_t$ and $f_{t_1}$ into (\ref{2-6})  and obtain
\begin{equation}\label{4-6}
A_t t_x+A_{t_1} f=\frac{\beta(f,n+1)}{t_1}-\frac{\beta(t_x,n)}{t}\, .
\end{equation}
The next system of two equations is the results of differentiation of (\ref{4-6}) with respect to $t_x$ consequently and usage of (\ref{beta6}) and (\ref{3-6}).
$$\left\{
\begin{array}{l}
\left(A_{t_1}+\frac{C}{t_1}\right)\beta(f, n+1)+\frac{1}{t_1}f=\frac{t_x}{t}+\left(\frac{C}{t}-A_t\right)\beta(t_x,n)\, ,\\
\left(CA_{t_1}+\frac{C^2-1}{t_1}\right)\beta(f, n+1)+\left(A_{t_1}+\frac{C}{t_1}\right)f=\left(\frac{C}{t}-A_t\right)t_x+\left(\frac{C^2-1}{t}-CA_t\right)\beta(t_x,n)\, .
\end{array}
\right.
$$
This system of two linear equations with respect to $\beta(f, n+1)$ and $f$ implies that $f$ can be written as
\begin{equation}\label{f-6}
f=\lambda(t,t_1)t_x+\mu(t,t_1)\beta(t_x, n)\, ,
\end{equation}
where
$$
\lambda=\frac{t_1-tt_1^2 A_t A_{t_1}-Ctt_1 A_t}{Ctt_1A_{t_1}+C^2t+t} \qquad {\mbox{and}}\qquad \mu=-\frac{t t_1 A_t +t_1^2A_{t_1}}{C t t_1 A_{t_1}+C^2t +t}\, .
$$
It follows from (\ref{f-6}) and (\ref{beta6}) that $f_{t_x}=\lambda+\mu\beta'(t_x, n)=\lambda-\mu\left(C+\frac{t_x}{\beta(t_x, n)}\right)$. On the other hand, by (\ref{3-6}), we have
$f_{t_x}=\beta(f, n+1)/\beta(t_x, n)$. Therefore,
\begin{equation}\label{betaf-6}
\beta(f, n+1)=-\mu t_x+(\lambda -C\mu )\beta(t_x, n)\, .
\end{equation}
We substitute $f=\lambda t_x+\mu \beta(t_x, n)$ and $\beta(f, n+1)=(\lambda -C\mu)\beta(t_x, n)-\mu t_x$ into (\ref{2-6}) and get
$$
\lambda_tt_x^2+\mu_t t_x\beta(t_x, n)+\lambda \lambda_{t_1} t_x^2+\lambda_{t_1}\mu \beta(t_x, n)t_x+\mu_{t_1}\lambda t_x\beta(t_x, n)+\mu_{t_1}\mu \beta^2(t_x, n)
$$
$$
=\frac{(\lambda -C\mu)^2\beta^2(t_x, n)}{t_1}-\frac{2(\lambda-C\mu)\mu t_x \beta(t_x, n)}{t_1}+\frac{\mu^2 t_x^2}{t_1}-\frac{(\lambda -C\mu)\beta^2(t_x, n)}{t}+\frac{\mu t_x\beta(t_x, n)}{t}\, ,
$$
that implies, after comparing coefficients before linearly independent functions $t_x^2$, $t_x\beta(t_x, n)$ and $\beta^2(t_x, n)$, the following system of equations on $\lambda$ and $\mu$ takes place
\begin{equation}\label{system-6}
\left\{
\begin{array}{l}
\lambda_t+\lambda \lambda_{t_1}=\frac{\mu^2}{t_1}\, ,\\
\mu_t+\lambda_{t_1}\mu+\mu_{t_1}\lambda =\frac{2(C\mu-\lambda)\mu}{t_1}+\frac{\mu}{t}\, ,\\
\mu_{t_1}\mu=\frac{(\lambda-C\mu)^2}{t_1}+\frac{C\mu -\lambda}{t}\, .
\end{array}
\right.
\end{equation}
Note that the Wronskian  of functions $t_x^2$, $t_x\beta(t_x, n)$ and $\beta^2(t_x, n)$ is equal to $2(t_x\beta'(t_x,n)-\beta(t_x, n))^3$. It is equal to $0$ if and only if $\beta(t_x, n)=\frac{-C\pm\sqrt{C^2-4}}{2}t_x$ provided that function $\beta$ satisfies (\ref{beta6}). In this case, due to (\ref{f-6}), we would have $t_{1x}=K(t, t_1)t_x$. Otherwise, the Wronskian is not $0$ that implies that functions $t_x^2$, $t_x\beta(t_x,n )$ and $\beta^2(t_x, n)$ are indeed linearly independent.

Let us find the relation between $\lambda$ and $\mu$. Denote by
$$
w=\frac{\beta(t_x, n)}{t_x}
\, .
$$
Equation (\ref{beta6}) becomes
\begin{equation}\label{ODE}
\frac{w \, dw}{w^2+Cw+1}=-\frac{dt_x}{t_x}\, .
\end{equation}
We study this equation in three different cases. \\
{\underline{Case 1)}} is when $C^2>4$ and,  therefore, $w^2+Cw+1=\left(w+\frac{C}{2}\right)^2-\frac{C^2-4}{4}$.\\
{\underline{Case 2)}} is when $C^2<4$ and,  therefore, $w^2+Cw+1=\left(w+\frac{C}{2}\right)^2+\frac{4-C^2}{4}$.\\
{\underline{Case 3)}} is when $C^2=4$ and,  therefore, $w^2+Cw+1=\left(w+\frac{C}{2}\right)^2$.\\

In Case 1) the solution of (\ref{ODE}) is
$$(w+B)^{-B^2}\left(w+\frac{1}{B}\right)t_x^{1-B^2}=Const_1(n)\,\qquad {\mbox{with}} \qquad B=\frac{C-\sqrt{C^2-4}}{2} \, ,$$
that can be rewritten as
\begin{equation}\label{Case1}
(\beta(t_x, n)+Bt_x)^{-B^2}(B\beta(t_x, n)+t_x)=Const_1\, .
\end{equation}
Also,
\begin{equation}\label{fCase1}
(\beta(f, n+1)+Bf)^{-B^2}(B\beta(f, n+1)+f)=Const_2\, .
\end{equation}
We substitute (\ref{betaf-6}) into (\ref{fCase1}), use (\ref{Case1}), and get that in Case 1) there is the following relation between $\lambda$ and $\mu$:
\begin{equation}\label{relationCase1}(B\lambda-\mu)^{-B^2}(\lambda-B\mu)=\nu(n), \quad  B=\frac{C-\sqrt{C^2-4}}{2}\, .
\end{equation}
Differentiation of (\ref{relationCase1}) with respect to $t$ and $t_1$ gives the following equations
\begin{equation}\label{mu}
\left\{
\begin{array}{l}
\mu\mu_t=(C\mu-\lambda)\lambda_t\, ,\\
\mu\mu_{t_1}=(C\mu-\lambda)\lambda_{t_1}\, .
\end{array}
\right.
\end{equation}

In Case 2) the solution of (\ref{ODE}) is
$$
\ln (w^2t_x^2+Cwt_x^2+t_x^2)-\frac{2C}{\sqrt{4-C^2}}\arctan\frac{2w+C}{\sqrt{4-C^2}}=Const_1\, ,
$$
that can be rewritten as
\begin{equation}\label{Case2}
\ln (\beta^2(t_x, n)+Ct_x\beta(t_x,n)+t_x^2)-\frac{2C}{\sqrt{4-C^2}}\arctan\frac{2\beta(t_x,n)+Ct_x}{t_x\sqrt{4-C^2}}=Const_1\, .
\end{equation}
Also,
\begin{equation}\label{fCase2}
\ln (\beta^2(f, n+1)+Cf\beta(f,n+1)+f^2)-\frac{2C}{\sqrt{4-C^2}}\arctan\frac{2\beta(f,n+1)+Cf}{f\sqrt{4-C^2}}=Const_2\, .
\end{equation}
We substitute (\ref{betaf-6}) into (\ref{fCase2}), use (\ref{Case2}), and get that in Case 2) there is the following relation between $\lambda$ and $\mu$:
\begin{equation}\label{relationCase2}\ln(\lambda^2-C\lambda \mu +\mu^2)-\frac{2C}{\sqrt{4-C^2}}\arctan \frac{2\lambda
-C\mu}{\mu\sqrt{4-C^2}}=\nu(n)\, .
\end{equation}
Differentiation of (\ref{relationCase2}) with respect to $t$ and $t_1$ gives (\ref{mu}).

We study Case 3) in the same way as Cases 1) and 2) and get  the following relation between $\lambda$ and $\mu$:
\begin{equation}\label{relationCase3}\left\{
\begin{array}{l}\ln(\lambda -\mu)+\frac{\mu}{\lambda -\mu}=\nu(n), \qquad {\mbox{if}} \qquad C=2,\\
\ln(\lambda +\mu)-\frac{\mu}{\lambda +\mu}=\nu(n), \qquad {\mbox{if}} \qquad C=-2.
\end{array}\right.
\end{equation}
Differentiation of (\ref{relationCase3}) with respect to $t$ and $t_1$ gives (\ref{mu}).

In all three cases we substitute the expressions for $\mu_t$ and $\mu_{t_1}$ from (\ref{mu}) into (\ref{system-6}) and have (\ref{lambda}). Note that system (\ref{lambda}) is compatible, i.e. $\lambda_{tt_1}=\lambda_{t_1t}$,
if and only if equations (\ref{mu}) hold.

\noindent {\bf{\underline{Finding $x$-integral}}}: Let us find function $F(t, t_1, t_2)$ such that $0=D_xF=F_t t_x+F_{t_1}t_{1x}+F_{t_2}t_{2x}$.
Due to (\ref{f-6}) and (\ref{betaf-6}), we have $t_{1x}=\lambda t_x+\mu\beta(t_x, n)$ and $t_{2x}=(\lambda_1\lambda-\mu_1\mu )t_x+(\lambda_1\mu+\mu_1\lambda-C\mu\mu_1)\beta(t_x,n)$, where $\lambda_1=D\lambda$ and $\mu_1=D\mu$.
By comparing the coefficients in $D_xF=0$ before $t_x$ and $\beta(t_x, n)$ we get the following system of two equations
$$
\left\{\begin{array}{l}
F_t+\lambda F_{t_1}+(\lambda_1\lambda-\mu_1\mu)F_{t_2}=0\, ,\\
\mu F_{t_1}+(\lambda_1\mu+\mu_1\lambda-C\mu\mu_1)F_{t_2}=0\, ,
\end{array}\right.
$$
that can be rewritten as
\begin{equation}\label{F}
\left\{\begin{array}{l}
\mu F_t+\mu_1(C\lambda \mu-\mu^2-\lambda^2)F_{t_2}=0\, ,\\
\mu F_{t_1}+(\lambda_1\mu+\mu_1\lambda-C\mu\mu_1)F_{t_2}=0\, .
\end{array}\right.
\end{equation}
Let $E(t, t_1, t_2)$ be such that $E_t=\frac{\mu^2+\lambda^2-C\lambda \mu}{\mu}$, $E_{t_1}=-\frac{\lambda_1}{\mu_1}-\frac{\lambda}{\mu}+C+\frac{1}{t_1}E$ and $E_{t_2}=\frac{1}{\mu_1}$. Such function $E$ exists since
$E_{tt_2}=0=E_{t_2 t}$ and $E_{tt_1}=E_{t_1 t}$, $E_{t_1t_2}=E_{t_2t_1}$ provided equations (\ref{mu}) hold.

In  new variables $\tilde{t}=E(t, t_1, t_2)$, $\tilde{t_1}=t_1$, $\tilde{t_2}=t_2$ the system (\ref{F}) becomes
\begin{equation}
\left\{\begin{array}{l}
F_{\tilde{t_2}}=0\, ,\\
\tilde{t}F_{\tilde{t}}+\tilde{t_1}F_{\tilde{t_1}}=0\, .
\end{array}\right.
\end{equation}
One can see that $x$-integral then can be taken as $F(t, t_1, t_2)=\frac{1}{t_1}E(t, t_1, t_2)$.

\section{Proof of Theorem \ref{newDiscretizationcase8}\label{THVIII}}

\noindent{\bf{\underline{Discretization}}}: Consider chains $t_{1x}=f(x,n,t,t_1,t_x)$ with $n$-integral
$I={\beta(t_x,n)}t_{xx}-\frac{1}{(x+\alpha(n))\beta(t_x,n)}$, where
\begin{equation}\label{beta8}
\beta'(t_x,n)=\beta^3(t_x,n)+\beta^2(t_x,n)\, .
\end{equation}
Denote by
$$
\beta=\beta(t_x,n)\, ,\quad \bar{\beta}=\beta(f,n+1)\, ,\quad \alpha=\alpha(n), \quad \alpha_1=\alpha(n+1)\, .
$$
Since $DI=I$ then
\begin{equation}\label{1-8}
\bar{\beta}(f_x+f_t t_x+f_{t_1}f+f_{t_x}t_{xx})-\frac{1}{(x+\alpha_1)\bar{\beta}}=\beta t_{xx}-\frac{1}{(x+\alpha)\beta}\, .
\end{equation}
By comparing the coefficients in (\ref{1-8}) before $t_{xx}$ we have
\begin{equation}\label{2-8}
\bar{\beta} f_{t_x}=\beta, \quad {\mbox{or}}\quad \gamma(f,n+1)-\gamma(t_x,n)=A(x,n,t,t_1) \quad {\mbox{with}} \quad \gamma'=\beta\, .
\end{equation}
It follows from (\ref{2-8}) that $f_t=A_t/\bar{\beta}$, $f_{t_1}=A_{t_1}/\bar{\beta}$ and $f_x=A_x/\bar{\beta}$. Substitute these expressions for $f_x$, $f_t$, $f_{t_1}$ into (\ref{1-8}) and get
\begin{equation} \label{3-8}
A_x+A_t t_x+A_{t_1}f=\frac{1}{x+\alpha_1}\bar{\mu}-\frac{1}{x+\alpha}\mu\, ,
\end{equation}
where
\begin{equation}\label{mu8}
\mu=\frac{1}{\beta(t_x,n)} \qquad {\mbox{and }} \qquad \bar{\mu} =\frac{1}{\beta(f, n+1)}\, .
\end{equation}
Note that equation (\ref{beta8}) in terms of $\mu$ can be rewritten as
\begin{equation}\label{mu8equation}-\mu \mu'=1+\mu\, .
\end{equation}
 Therefore,
\begin{equation}\label{mu8solution}
\mu(t_x,n)-\ln (1+\mu(t_x,n)) +t_x=C_1\, ,
\end{equation}
where $C_1$ is some constant depending on $n$ only.\\
We differentiate (\ref{3-8}) with respect to $t_x$, use (\ref{2-8}), (\ref{mu8}), (\ref{mu8equation}) and get
\begin{equation}\label{4-8}
A_t+A_{t_1}\frac{\bar{\mu}}{\mu}=-\frac{1}{x+\alpha_1}\frac{1+\bar{\mu}}{\mu}+\frac{1}{x+\alpha}\frac{1+\mu}{\mu}\, ,
\end{equation}
that is equivalent to
\begin{equation}
\label{mu8f}
\mu(f, n+1)=\frac{(x+\alpha)^{-1}-A_t}{(x+\alpha_1)^{-1}+A_{t_1}}\mu(t_x,n)+\frac{(x+\alpha)^{-1}-(x+\alpha_1)^{-1}}{A_{t_1}+(x+\alpha_1)^{-1}}\, ,
\end{equation}
or $A_{t_1}=-(x+\alpha_1)^{-1}$, $A_t=(x+\alpha)^{-1}$ and $\alpha=\alpha_1$ (in this case $\gamma(t_{1x})=\gamma(t_x)+(t-t_1)(x+\alpha)^{-1}$ by (\ref{2-8})).\\
Differentiate (\ref{4-8}) with respect to $t_x$, use (\ref{2-8}), (\ref{mu8}) and (\ref{mu8equation}), and get
$$
A_{t_1}(\bar{\mu}-\mu)+\frac{1+\bar{\mu}}{x+\alpha_1}=\frac{1+\mu}{x+\alpha}\, ,
$$
or
\begin{equation}\label{muf8}
\mu(f,n+1)=\left(\frac{A_{t_1}+(x+\alpha)^{-1}}{A_{t_1}+(x+\alpha_1)^{-1}}\right)\mu(t_x,n)+\frac{(x+\alpha)^{-1}-(x+\alpha_1)^{-1}}{A_{t_1}+(x+\alpha_1)^{-1}}\, .
\end{equation}
By comparing the last equation with (\ref{mu8f}) we get
\begin{equation}\label{A8}
A_t+A_{t_1}=0\, .
\end{equation}
Note that, by (\ref{muf8}), we have
\begin{equation}\label{5-8}
1+\bar{\mu}=\frac{A_{t_1}+(x+\alpha)^{-1}}{A_{t_1}+(x+\alpha_1)^{-1}}(1+\mu)\, .
\end{equation}
It follows from (\ref{mu8solution}) that
\begin{equation}\label{6-8}
\mu(f,n+1)-\ln (1+\mu(f,n+1))+f=C_2\, .
\end{equation}
Substitute (\ref{5-8}), (\ref{muf8}) into (\ref{6-8}), use (\ref{mu8solution}) and obtain
\begin{equation}\label{f8}
f=(1-K)\mu(t_x,n)+t_x+(\ln K -K)+C_2-C_1+1, \quad
{\mbox{where}}\quad
K=\frac{A_{t_1}+(x+\alpha)^{-1}}{A_{t_1}+(x+\alpha_1)^{-1}}\, .
\end{equation}
Observe that
$$K_t=\left(1+\frac{(x+\alpha)^{-1}-(x+\alpha_1)^{-1}}{A_{t_1}+(x+\alpha_1)^{-1}}\right)_t=
\frac{(x+\alpha_1)^{-1}-(x+\alpha)^{-1}}{(A_{t_1}+(x+\alpha_1)^{-1})^2}A_{t_1t}
$$
$$=\frac{(x+\alpha_1)^{-1}-(x+\alpha)^{-1}}{(A_{t_1}+(x+\alpha_1)^{-1})^2}A_{tt_1}=-\frac{(x+\alpha_1)^{-1}-(x+\alpha)^{-1}}{(A_{t_1}+(x+\alpha_1)^{-1})^2}A_{t_1t_1}=-K_{t_1}\,
$$
by (\ref{A8}), i.e.
\begin{equation}\label{partialK}
K_t+K_{t_1}=0\, .
\end{equation}
Substitute (\ref{f8}) into (\ref{1-8}), use (\ref{muf8}) and (\ref{partialK}), compare the coefficients in the obtained equality before linearly independent functions $t_x^0$, $\mu$ and $\mu^2$ (the Wronskian of $t_x^0$, $\mu$ and $\mu^2$ is equal to $-2(1+\mu)^3\mu^{-3}\ne0 $ unless $\mu=-1$), and get
that function $K(x,n,t,t_1)$ must satisfy (\ref{functionK}). One can check that system (\ref{functionK}) is consistent since $(K-1)^2(\ln K)_{t_1 x}=(K-1)^2(\ln K)_{x t_1}$, i.e. $K_{t_1 x}=K_{x t_1}$.

\noindent{\bf{\underline{Finding $x$-integral}}}: Let us find function $F(x,t,t_1,t_2)$ such that
$$
0=D_xF=F_x+F_t t_x+F_{t_1}t_{1x}+F_{t_2}t_{2x}\, .
$$
Note that, due to the fact that $t_{1x}=(1-K)\mu(t_x, n) +t_x+(-K+\ln K)$ and $\mu(f, n+1)=K\mu(t_x,n)+(K-1)$ by (\ref{muf8}), we have
$$
t_{2x}=(1-KK_1)\mu(t_x,n)+t_x+(-1-KK_1+\ln(KK_1))\, .
$$
Functions $t_x^0$, $t_x$ and $\mu(t_x, n)$ are linearly independent since their Wronskian is equal to $-(1+\mu)\mu^{-3}\ne 0$ unless $\mu=-1$.
We compare the coefficients before  $t_x^0$, $t_x$ and $\mu(t_x, n)$  in $D_x F=0$ and get
\begin{equation}\label{x8}\left\{\begin{array}{l}
F_x+(\ln K-K)F_{t_1}+(\ln (KK_1)-(KK_1)-1)F_{t_2}=0\, ,\\
F_t+F_{t_1}+F_{t_2}=0\, ,\\
(1-K)F_{t_1}+(1-KK_1)F_{t_2}=0\, .
\end{array}\right.
\end{equation}
In new variables $\tau=t$, $\tau_1=t_1-t$ and $\tau_2=t_2-t_1$ the system (\ref{x8}) can be written as
\begin{equation}\label{x8new}\left\{\begin{array}{l}
A(F)=F_x+\left\{\frac{K(1-K_1)(1-\ln K)}{1-K}+\ln K_1 -1\right\}F_{\tau_2}=0\, ,\\
F_{\tau}=0\, ,\\
B(F)=(1-K)F_{\tau_1}+K(1-K_1)F_{\tau_2}=0\, .
\end{array}\right.
\end{equation}
One can check that the last system is closed since $[A,B]=(1-K^{-1})K_{\tau_1}A+K_x K^{-2}B$. Note that $K=K(\tau_1)$ and then $K_1=K(\tau_2)$.
 Define function $ E(x,t,t_1,t_2)$, where
$E_x=\frac{K(1-\ln K)}{1-K}-\frac{1-\ln K_1}{1-K_1}+\frac{1}{x+\alpha(n+1)}E$, $E_{\tau_1}=\frac{K}{1-K}$, $E_{\tau_2}=-\frac{1}{1-K_1}$. Such function exists
since $E_{\tau_1\tau_2}=0=E_{\tau_2\tau_1}$ and $E_{x \tau_1}=E_{\tau_1 x}$, $E_{x \tau_2}=E_{\tau_2 x}$ due to (\ref{functionK}) and the fact that $K_{\tau_1}(\tau_1)=K_{t_1}(t,t_1)$.\\
Introduce $\tau_1^*=\tau_1$ and $\tau_2^*=E(x,\tau_1, \tau_2)$. The first and the third equations of (\ref{x8new}) become
\begin{equation}\left\{\begin{array}{l}
F_x+\frac{\tau_2^*}{x+\alpha_1}F_{\tau_2^*}=0\, ,\\
F_{\tau_1^*}=0\,  ,
\end{array}\right.
\end{equation}
that implies that $x$-integral can be taken as $F=\frac{1}{x+\alpha_1}E(x, \tau_1, \tau_2)$.

\section{Proof of Theorem \ref{newDiscretization}\label{THIV,VII}}

\noindent{\bf{\underline{Discretization, Part(a)}}}: We consider semi-discrete equations $t_{1x}=f(x,n,t,t_1, t_x)$ with  $n$-integral
\begin{equation}\label{case4n-integral}
I=\frac{t_{xx}}{t_x}-\frac{2t_x}{t-x}+\frac{1}{t-x}
\end{equation}
From $DI=I$ we get
\begin{equation}\label{case4(1)}
\frac{f_x+f_t t_x+f_{t_1}f+f_{t_x}t_{xx}}{f}-\frac{2f}{t_1-x}+\frac{1}{t_1-x}=\frac{t_{xx}}{t_x}-
\frac{2t_x}{t-x}+\frac{1}{t-x}
\end{equation}
By comparing the coefficients in (\ref{case4(1)}) before $t_{xx}$ we obtain $f_{t_x}/f=1/t_x$, or
$f=t_xK$, where $K$ is some function depending on $x$, $n$, $t$ and $t_1$. Substitute $f=t_xK$ into
(\ref{case4(1)}) and find
\begin{equation}\label{case4(2)}
\frac{K_xt_x+K_tt_x^2+K_{t_1}Kt_x^2}{Kt_x}-\frac{2Kt_x}{t_1-x}+\frac{1}{t_1-x}=-\frac{2t_x}{t-x}+\frac{1}{t-x}
\end{equation}
Compare the coefficients before $t_x$ and $t_x^0$ in (\ref{case4(2)}) and get
\begin{equation}\label{case4(3)}
\frac{K_t}{K}+K_{t_1}=\frac{2K}{t_1-x}-\frac{2}{t-x}
\end{equation}
\begin{equation}\label{case4(4)}
\frac{K_x}{K}=\frac{1}{t-x}-\frac{1}{t_1-x}
\end{equation}
We solve (\ref{case4(4)}) and have $K=C(t_1-x)/(t-x)$, where $C$ is some function depending on $n$, $t$ and $t_1$.
Substitute this expression for $K$ into (\ref{case4(3)}) and obtain
\begin{equation}\label{case4(5)}
\frac{C_t}{C}(t-x)+C_{t_1}(t_1-x)=C-1
\end{equation}
By comparing the coefficients before $x$ and $x^0$ in (\ref{case4(5)}) we get the system of equations
$$
\left\{
\begin{array}{l}
\displaystyle{\frac{C_t}{C}+C_{t_1}=0}, \\
\\
\displaystyle{\frac{C_t}{C}t+C_{t_1}t_1=C-1}
\end{array}
\right.
$$
whose solution is $C=(1+M(n)t_1)/(1+M(n)t)$. Thus, equation $t_{1x}=f(x,n,t,t_1, t_x)$ possessing $n$-integral
(\ref{case4n-integral})
is
\begin{equation}\label{case4discretization}
t_{1x}=\frac{(1+M(n)t_1)(t_1-x)}{(1+M(n)t)(t-x)}t_x,
\end{equation}
where $M(n)$ is an arbitrary function depending on $n$ only.

\noindent{\bf{\underline{Finding $x$-integral, Part(a)}}}: Let us find an $x$-integral of equation  (\ref{case4discretization}) of minimal order if it exists.
First, assume that equation (\ref{case4discretization}) possesses an $x$-integral $F(x,n,t, t_1)$
of the first order.  The equality $D_xF(x, n, t, t_1)=0$ can be rewritten as
\begin{equation} \label{case4x-integralTheFirst}
\displaystyle{F_x+F_tt_x  +F_{t_1} \frac{(1+M(n)t_1)(t_1-x)}{(1+M(n)t)(t-x)}t_x=0}
\end{equation}
By comparing the coefficients before $t_x^0$ and $t_x$ we get
\begin{equation}\label{case4x-integral(1)TheFirst}
F_x=0
\end{equation}
and
\begin{equation}\label{case4x-integral(2)TheFirst}
F_t  +F_{t_1} \frac{(1+M(n)t_1)(t_1-x)}{(1+M(n)t)(t-x)}+
=0
\end{equation}
We differentiate equation (\ref{case4x-integral(2)TheFirst}) with respect to $x$,
use   (\ref{case4x-integral(1)TheFirst}), and get a contradictory equality
$$
\frac{\partial }{ \partial x}\left\{ \frac{t_1 -x}{t-x} \right\}=0\, .
$$
It means that equation (\ref{case4discretization}) does not possess an $x$-integral $F(x,n,t, t_1)$
of the first order.

Now let us see whether equation (\ref{case4discretization}) possesses an $x$-integral $F(x,n,t, t_1, t_2)$
of the second order.
Since $D_xF=0$ then
\begin{equation} \label{case4x-integral}
\begin{array}{ll}
\displaystyle{F_x+F_tt_x  +F_{t_1} \frac{(1+M(n)t_1)(t_1-x)}{(1+M(n)t)(t-x)}t_x}\\
\\
\displaystyle{+F_{t_2}\frac{(1+M(n+1)t_2)(t_2-x)(1+M(n)t_1)(t_1-x)}{(1+M(n+1)t_1)(t_1-x)(1+M(n)t)(t-x)}t_x=0}
\end{array}
\end{equation}
By comparing the coefficients before $t_x^0$ and $t_x$ we get
\begin{equation}\label{case4x-integral(1)}
F_x=0
\end{equation}
and
\begin{equation}\label{case4x-integral(2)}
F_t  +F_{t_1} \frac{(1+M(n)t_1)(t_1-x)}{(1+M(n)t)(t-x)}+
F_{t_2}\frac{(1+M(n+1)t_2)(t_2-x)(1+M(n)t_1)}{(1+M(n+1)t_1)(1+M(n)t)(t-x)}=0
\end{equation}
We differentiate equation (\ref{case4x-integral(2)}) with respect to $x$ and get
\begin{equation}\label{case4x-integral(3)}
F_{t_1} (t_1-t)+
F_{t_2}\frac{(1+M(n+1)t_2)(t_2-t)}{(1+M(n+1)t_1)}=0
\end{equation}
One can check that the system of partial differential equations
(\ref{case4x-integral(1)}), (\ref{case4x-integral(2)}) and  (\ref{case4x-integral(3)})
is closed. To solve this system of equations we use the famous Jacobi Method:
we first diagonalise the system (that is, we make it normal) and then we do the
necessary changes of variables
using the first integrals of the equations from the system. The calculations are standard but rather long.
That is why we omit these straightforward steps and present an $x$-integral immediately. It is
\begin{equation} \label{case4x-integralFinal}
F(x,n, t,t_1,t_2)=\frac{(1+M(n+1)t_2)(t_1-t)}{(1+M(n)t)(t_1-t_2)}
\end{equation}

For the readers familiar with the characteristic rings (see \cite{ZhiberBook}, \cite{HZhP2008}, \cite{HZhP2009}) we would like to  note that the existence of a
nontrivial $x$-integral for equation (\ref{case4discretization}) implies
that the characteristic ring $L_x$ in $x$-direction for this equation is of finite dimension.
 It is not difficult to see that for equation (\ref{case4discretization})
characteristic ring  $L_x$ is generated by three vector fields

\begin{equation}
\begin{array}{l}
\displaystyle{X_1=\frac{\partial}{\partial x}\, ,}\\
\\
\displaystyle{X_2=\frac{\partial }{\partial t }  + \frac{(1+M(n)t_1)(t_1-x)}{(1+M(n)t)(t-x)}\frac{\partial }{\partial t_1 } +
\frac{(1+M(n+1)t_2)(t_2-x)(1+M(n)t_1)}{(1+M(n+1)t_1)(1+M(n)t)(t-x)}\frac{\partial }{\partial t_2} \, ,}\\
\\
\displaystyle{X_3= (t_1-t)\frac{\partial }{\partial t_1}+
\frac{(1+M(n+1)t_2)(t_2-t)}{(1+M(n+1)t_1)}\frac{\partial }{\partial t_2 }\, .}
\end{array}
\end{equation}
In partiacular, it means that the dimension of $L_x$ for equation (\ref{case4discretization})
 is $3$.

\noindent{\bf{\underline{Discretization, Part (b)}}}:  Let us consider semi-discrete equations (\ref{dhyp}) possessing $n$-integral
\begin{equation}\label{case7n-integral}
\displaystyle{I=\frac{t_{xx}}{\sqrt{t_x}}+\frac{2\sqrt{t_x}}{x+\varepsilon n}}\end{equation}
Since $DI=I$  then
\begin{equation}\label{case7}
\frac{f_x+f_t t_x +f_{t_1 }f +f_{t_x}t_{xx}}{\sqrt{f}}+\frac{2\sqrt{f}}{x+\varepsilon (n+1)}=\frac{t_{xx}}{\sqrt{t_x}}
+\frac{2\sqrt{t_x}}{x+\varepsilon n}
\end{equation}
We compare the coefficients before $t_{xx}$ in (\ref{case7})   and get
$f_{t_x}/\sqrt{f}=1/\sqrt{t_x}$, or $\sqrt{f}=\sqrt{t_x}+L$, where $L$
is some function depending on $x$, $n$, $t$, $t_1$. We substitute $f=(\sqrt{t_x}+L)^2$ into (\ref{case7})
and have
$$
L_x+L_tt_x+L_{t_1}(\sqrt{t_x}+L)^2+\frac{\sqrt{t_x}+L}{x+\varepsilon (n+1)}=\frac{\sqrt{t_x}}{x+\varepsilon n}\,
$$
that implies that function $L(x,n, t, t_1)$ satisfies the following three differential equations
\begin{equation}\label{case7(1)}
L_t+L_{t_1}=0
\end{equation}
\begin{equation}\label{case7(2)}
2LL_{t_1}+\frac{1}{x+\varepsilon (n+1)}=\frac{1}{x+\varepsilon n}
\end{equation}
\begin{equation}\label{case7(3)}
L_x +L^2L_{t_1}+\frac{L}{x+\varepsilon (n+1)}=0
\end{equation}
Equation (\ref{case7(2)}) gives that
\begin{equation}\label{case7(4)}
L^2=\left(\frac{1}{x+\varepsilon n}-\frac{1}{x+\varepsilon (n+1)} \right)t_1+M\,
\end{equation}
where $M$ is some function depending on $x$, $n$ and $t$.
We substitute the expression for $L^2$ from (\ref{case7(4)}) into the equation (\ref{case7(1)})
rewritten as $LL_t+LL_{t_1}=0$ and obtain
$$
M=\left(\frac{1}{x+\varepsilon (n+1)}-\frac{1}{x+\varepsilon n}\right)t+K\,
$$
where $K$ is some function depending on $x$ and $n$ only.
Thus,
$$
L^2=\left(\frac{1}{x+\varepsilon n}-\frac{1}{x+\varepsilon (n+1)}\right)(t_1-t)+K
$$
Substitute this expression for $L^2$ into the equation  (\ref{case7(3)}) multiplied by $2L$
and have
$$
K_x=\left(\frac{1}{x+\varepsilon (n+1)}-\frac{1}{x+\varepsilon n}\right)K
\qquad \to\qquad
K=\frac{C(n)}{(x+\varepsilon n)(x+\varepsilon (n+1))},$$
where $C(n)$ is an arbitrary function of $n$. Therefore,
$$
L^2=\frac{\varepsilon (t_1-t)+C(n)}{(x+\varepsilon n)(x+\varepsilon (n+1))}\,
$$
and then
\begin{equation}\label{case7disretizationMiddle}
f(x,n,t,t_1,t_x)=\left(\sqrt{t_x}+\sqrt{\frac{\varepsilon (t_1-t)+C(n)}{(x+\varepsilon n)(x+\varepsilon (n+1))}}\right)^2
\end{equation}
Let us note that one can eliminate function $C(n)$ in (\ref{case7disretizationMiddle}) by the change of variable
$t_(x,n)=\tau(x,n)+d(n)$, where $d(n)$ satisfies $\varepsilon (d(n+1)-d(n)) +C(n)=0$.
Equations possessing $n$-integral (\ref{case7n-integral}) become \begin{equation}\label{case7disretization}
t_{1x}=\left(\sqrt{t_x}+\sqrt{\frac{\varepsilon (t_1-t)}{(x+\varepsilon n)(x+\varepsilon (n+1))}}\right)^2
\end{equation}

\noindent{\bf{\underline{Finding $x$-integral, Part (b)}}}:  Let us find $x$-integral of equation (\ref{case7disretization}).
Denote by
\begin{equation}\label{alpha-beta}
\alpha =\sqrt{\frac{\varepsilon (t_1-t)}{(x+\varepsilon n)(x+\varepsilon (n+1))}}\,   \qquad
\beta =D\alpha=\sqrt{\frac{\varepsilon (t_2-t_1)}{(x+\varepsilon (n+1))(x+\varepsilon (n+2))}}
\end{equation}

We find an $x$-integral of the minimal order of equation  (\ref{case7disretization})
in the same way as we did for equation (\ref{case4discretization}).
We look for function $F(x,n,t,t_1,t_2)$ such that $D_xF=0$. We have,
\begin{equation}\label{x-integral}
F_x+F_tt_x+F_{t_1}(t_x+\alpha^2+2\sqrt{t_x}\alpha)+F_{t_2}(\sqrt{t_x}+\alpha +\beta)^2=0
\end{equation}
Compare the coefficients before $t_x$, $\sqrt{t_x}$ and $t_x^0$ in (\ref{x-integral}) and get
the following system of equation
$$\left\{\begin{array}{l}
F_t+F_{t_1}+F_{t_2}=0\,\\
\alpha F_{t_1}+(\alpha +\beta)F_{t_2}=0\, \\
F_x+\alpha^2F_{t_1}+(\alpha +\beta)^2 F_{t_2}=0
\end{array}\right.
$$
that can be rewritten as
$$\left\{\begin{array}{l}
F_x+\beta(\alpha+\beta)F_{t_2}=0\,\\
\alpha F_t-\beta F_{t_2 =0}\, \\
\alpha F_{t_1}+(\alpha +\beta)F_{t_2}=0
\end{array}\right.
$$
One can check that the system is closed and its solution is
\begin{equation}\label{x-integral-function}
F=(x+\varepsilon n)\alpha -(x+\varepsilon (n+2))\beta\,  .
\end{equation}

\section{Continuum limits. Proof of Theorem \ref{ContinuumLimits}}\label{C}

\noindent {\bf{\underline{ Case $F$}}}:
In semi-discrete equation (\ref{case4discretization}) we rewrite
$t(x,n)$ as $u(x,y)$,  $t_1$ as $u(x,y)+\varepsilon u_y(x,y)$, $M(n)$ as $1/R(\varepsilon n)=1/R(y)$,
and  get
$$
u_x+\varepsilon u_{xy}=\left(\frac{R(y)+u+\varepsilon u_y}{R(y)+u}\right)
\left(\frac{u+\varepsilon u_y-x}{u-x}\right)u_x\, ,
$$
or
$$
u_x+\varepsilon u_{xy}=\left( 1+\frac{\varepsilon u_y}{u+R(y)}\right)
\left(1+\frac{\varepsilon u_y}{u-x}\right) u_x\, ,
$$
or
$$
u_{xy}=u_xu_y\left( \frac{1}{u-x}+\frac{1}{u+R(y)}\right) +
\varepsilon \frac{ u_y^2u_x}{(u-x)(u+R(y))}
$$
Now we  let $\varepsilon$ approach $0$ to get
continuous   equation analogue
\begin{equation}\label{case4continuous}
u_{xy}=\left(\frac{1}{u-x}+\frac{1}{u+R(y)}\right)u_xu_y\, .
\end{equation}
Note that after the change of variable $\tilde{y}=-R(y)$ equation (\ref{case4continuous}) becomes
 \begin{equation}\label{case4continuousLiouville}
u_{x\tilde{y}}=\left(\frac{1}{u-x}+\frac{1}{u-\tilde{y}}\right)u_xu_{\tilde{y}}\, .
\end{equation}
In $x$-integral $\varepsilon^{-1}(1+(1+n^{-1})F)$ of semi-discrete equation (\ref{case4discretization}),
where $F$ is taken as  (\ref{case4x-integralFinal})  we substitute
$u$, $u+\varepsilon u_y+(1/2)\varepsilon^2u_{yy}$, $1/R(y)$ and $y$ instead of $t$, $t_1$, $M(n)$
and $\varepsilon n$
correspondingly, and let
$\varepsilon $ approach $0$ to get
 its continuous analogue
\begin{equation}\label{x-intcase4continuous}
\tilde{F}=-\frac{u_{yy}}{{u_y}}+\frac{R'(y)}{u+R(y)}+\frac{2u_y}{u+R(y)}
\end{equation}
Note that continuous equation (\ref{case4continuous}) possesses $y$-integral
(\ref{case4n-integral}) and $x$-integral (\ref{x-intcase4continuous})

\noindent {\bf{\underline{ Case $G$}}}:
In semi-discrete equation (\ref{case7disretization}) we substitute
$u$, $u+\varepsilon u_y$ and $y$ instead of $t$, $t_1$ and $\varepsilon n$ correspondingly, and let
$\varepsilon $ approach $0$ to get its
continuous  Liouville equation analogue
\begin{equation}\label{case7continuous}
u_{xy}=\frac{2\sqrt{u_{x}u_y}}{x+y}
\end{equation}
In $x$-integral (\ref{x-integral-function}) multiplied by $-2\varepsilon^{-2}$ we substitute
$u$, $u+\varepsilon u_y+(1/2)\varepsilon^2u_{yy}$ and $y$ instead of $t$, $t_1$ and $\varepsilon n$
correspondingly, and let
$\varepsilon $ approach $0$ to get
 its continuous analogue
\begin{equation}\label{x-intcase7continuous}
\tilde{F}=\frac{u_{yy}}{\sqrt{u_y}}+\frac{2}{x+y}
\end{equation}
Note that continuous equation (\ref{case7continuous}) possesses $y$-integral
$$
\displaystyle{I=\frac{u_{xx}}{\sqrt{u_x}}+\frac{2\sqrt{u_x}}{x+y}}$$
which is a continuous analogue of (\ref{case7n-integral}) and $x$-integral (\ref{x-intcase7continuous})

\section{Discretization and the  B\"{a}cklund  Transformation}\label{BacklandDiscretization}

Recall the definition of the  B\"{a}cklund  transformation for the PDE (see \cite{Newell}). Suppose that $u(x,t)$ and $\tilde{u}(x,t)$ satisfy respectively differential equations
\begin{equation}\label{*1}
E[u]=0
\end{equation}
and
\begin{equation} \label{*2}
\tilde{E}[\tilde{u}]=0\, .
\end{equation}
Here the expression $E[u]$ denotes the fact that $E$ depends on $u$ and a finite number of its derivatives. Then the set of the relations \begin{equation}\label{*3} R_j[u,\tilde{u}]=0,\qquad j=1,2,\ldots, k\end{equation} defines the  B\"{a}cklund transformation if these relations satisfy the following conditions: $\tilde{u}$ exists and solves (\ref{*2}) whenever $u$ exists and solves (\ref{*1}) and vice versa. When $u$ and $\tilde{u}$ are solutions of one and the equation then (\ref{*3}) defines the  B\"{a}cklund autotransformation. In that case we exclude the trivial autotransformation $u\equiv\tilde{u}$.

It is well-known that iterations of the  B\"{a}cklund autotransformation of a PDE define a semidiscrete equation. Semi-discrete models constructed in such a way are also called discretizations. Below we examine the question whether the semi-discrete equations found above by discretization preserving integrals do realize the  B\"{a}cklund autotransformation. The answer is stated in the following proposition.                                                                                                                                           \begin{pro}\label{ProBacklund}
In cases $A$, $B$, $C$, $D$, $E$, $F$ from Theorem \ref{ContinuumLimits} the semi-discrete equations realize the  B\"{a}cklund autotransformations for their continuum limits,
but in the case $G$ does not.
\end{pro}
{\it{Scheme of the proof}}. For the case $B$ the proof is very simple.
By differentiation of the equation
\begin{equation}\label{*4} u_{1x}=u_x-e^u+e^{u_1}\end{equation}
with respect to $y$ we find the equation
\begin{equation}\label{*5} u_{1xy}-e^{u_1}u_{1y}=u_{xy}-e^uu_y\,\end{equation}
which is satisfied identically by means of the equation (II) from the Goursat list: $u_{xy}=e^uu_y$.
Equation (\ref{*5}) immediately shows that all requests of the definition of the   B\"{a}cklund transformation are satisfied.

 Concentrate on the case $A$ :
 \begin{equation}\label{*6}u_{1x}=u_x+Ce^{(u_1+u)/2}\end{equation}
 which is a discretization of the Liouville equation
 \begin{equation} \label{*7}u_{xy}=e^u \end{equation}
 Differentiate (\ref{*6}) with respect to $y$  and get
 \begin{equation}\label{*8}
 u_{1xy}=e^u+(1/2)Ce^{(u_1+u)/2}(u_{1y}+u_y)
\end{equation}
By differentiating (\ref{*8}) with respect to $x$ and simplifying by means of (\ref{*6})-(\ref{*8}) we get
\begin{equation}\label{*9} u_{1xxy}-u_{1xy}u_{1x}=0.\end{equation}
Reduce it to the convenient form $d(-u_1+\log u_{1xy})/dx=0$ and then integrate
\begin{equation}\label{*10}
u_{1xy}=C_1(y)e^{u_1}\, .
\end{equation}
Due to (\ref{*10}) equation (\ref{*8}) is rewritten as
\begin{equation}\label{*11}
u_{1y}=-u_y+C_1(y) e^{(u_1-u)/2}-e^{(u-u_1)/2}\, .
\end{equation}
Reasonings above result in the statement: relations (\ref{*6}), (\ref{*11}) define the  B\"{a}cklund transformation between equations (\ref{*7}) and (\ref{*10}). Choose $C_1(y)=1$ then this transformation becomes the B\"{a}cklund auto-transformation  for the Liouville equation which has been found by A.V.B\"{a}cklund
himself (see \cite{IbragimovBook}).

Consider the case $G$. Let us prove that
\begin{equation}\label{*12}
\sqrt{u_{1x}}=\sqrt{u_x}+\sqrt{\frac{\varepsilon(u_1-u)}{(x+\varepsilon n)(x+\varepsilon (n+1))}}
\end{equation}
does not realize the B\"{a}cklund autotransformation for the equation
\begin{equation}\label{*13}
u_{xy}=\frac{2\sqrt{u_x}\sqrt{u_y}}{x+y}\, .
\end{equation}
Assume  contrary and differentiate (\ref{*12}) with respect to $y$. After simplification we get
\begin{equation}\label{*14}
\sqrt{u_{1y}}+\sqrt{u_y} =\frac{2\sqrt{x+\varepsilon n}\sqrt{x+\varepsilon(n+1)}}{(x+y)\sqrt{\varepsilon}}\sqrt{u_1-u}\, .
\end{equation}
Now differentiate (\ref{*14}) with respect to $x$ and simplify by means of (\ref{*12})-(\ref{*14}). As a result one gets a contradictory equation
$$ \frac{\sqrt{u_1-u}}{\sqrt{\varepsilon}}
\left(
\frac{\sqrt{x+\varepsilon (n+1)}}{\sqrt{x+\varepsilon n}}+
\frac{\sqrt{x+\varepsilon n}}
{\sqrt{x+\varepsilon(n+1)}}-2\frac{\sqrt{x+\varepsilon n}\sqrt{x+\varepsilon(n+1)}}{x+y}\right)=0\, .$$
This proves that (\ref{*12}) does not realize the B\"{a}cklund autotransformation for (\ref{*13}). Other statements of the Proposition \ref{ProBacklund} are proved in a similar way.

\section{Conclusion}
Darboux integrable equations or equations of Liouville type constitute a very well studied subclass of hyperbolic type PDE. The problem of complete description of this subclass was formulated and partly solved by E.Goursat in 1899 (see \cite{Goursat}). Since then many authors have investigated the classification problem \cite{ZhiberBook}-\cite{Kaptsov}. To the best of our knowledge the problem up to now is still unsolved. The similar problem for the semi-discrete chains (\ref{dhyp}) and the fully discrete models is less studied. We can mention only particular classes of the equations investigated in  \cite{HZhP2008} - \cite{Yamilov}, \cite{ZhZh1} and  \cite{ZhZh2}. In the present article we discussed the problem of discretization via integrals and presented some new non-autonomous examples of the Darboux integrable chains.

\section*{Acknowledgments}
This work is partially supported by Russian Foundation for
Basic Research (RFBR) grant 14-01-97008-r-povolzhie-a.


\begin{thebibliography}{EMG}

\bibitem[1]{Ragnisco}M. Bruschi, D. Levi, and  O. Ragnisco, {\it  Discrete version of the nonlinear
Schrödinger equation with linearlyx-dependent coefficients}, Il Nuovo Cimento A Series 11, 53(1),
21-30 (1979).

\bibitem[2]{Suris} Y. B. Suris, {\it The problem of integrable discretization: Hamiltonian approach},
 Vol. 219. Springer, 2003.


\bibitem[3]{Hirota} R. Hirota and K. Kimura, {\it Discretization of the Euler top},
Journal of the Physical Society of Japan 69 (2000): 627.

\bibitem[4]{Murata} M. Murata, et al.,{\it  How to discretize differential systems in a
systematic way},  Journal of Physics A: Mathematical and Theoretical 43.31 (2010): 315203.

\bibitem[5]{Veselov} J. Moser and A. P. Veselov, {\it Discrete versions of some classical
integrable systems and factorization of matrix polynomials}, Communications in Mathematical
Physics 139.2 (1991): 217-243.

\bibitem[6]{Gibbons} J. Gibbons  and B. A. Kupershmidt, {\it Time discretizations of lattice
integrable systems}, Physics Letters A 165.2 (1992): 105-110.

\bibitem[7]{Dmitry} D. Zakharov, {\it A discrete analogue of the modified Novikov-Veselov hierarchy},
  arXiv.org.nlin.arXiv:0904.3728v2

\bibitem[8]{Adler} V. E. Adler,{\it On a discrete analog of the Tzitzeica equation}, (arXiv:1103.5139)

\bibitem[9]{Discretization} I. T. Habibullin, N. Zheltukhina, and A. Sakieva,
{\it Discretization of hyperbolic type Darboux integrable equations preserving integrability},
J. Math. Phys. 52 (2011), 093507.

\bibitem[10]{Goursat} E. Goursat,  {\it Recherches sur quelques $\acute{e}$quations aux d$\acute{e}$riv$\acute{e}$s
partielles du
second
ordre}, Annales de la facult$\acute{e}$ des Sciences de l'Universit$\acute{e}$ de Toulouse 2e
s$\acute{e}$rie, 1:1
(1899), 31-78.

\bibitem[11]  {HZhP2008} I. Habibullin I, N. Zheltukhina and A. Pekcan, {\it  On the classification of
Darboux integrable chains}, J. Math.
Phys. 49, 2008, 102702

\bibitem[12]  {HZhP2009} I.  Habibullin, N. Zheltukhina and A. Pekcan, {\it Complete list
of Darboux integrable chains of the form
$t_{1x} = t_x + d(t, t_1)$}, J. Math. Phys. 50, 2009 102710


\bibitem[13]{Yamilov} R. N. Garifullin and R. I. Yamilov, {\it Generalized symmetry classification of
discrete equations of a class depending on twelve parameters}, Journal of Physics A:
Mathematical and Theoretical 45.34 (2012): 345205.


\bibitem[14]  {Newell} Alan C. Newell,  {\it Solitons in mathematics and physics},
Philadelphia: Society for Industrial and applied Mathematics, 1985.

\bibitem[15]  {IbragimovBook}  Nail H. Ibragimov, {\it Transformation groups applied to
mathematical physics}, Vol. 3. Springer, 2001.


\bibitem[16]  {ZhiberBook} A. V. Zhiber, R.D. Murtazina, I. T Habibullin and  A. B. Shabat,
{\it Characteristic Lie rings and nonlinear integrable equations},-- M.-Izhevsk, 2012. -- pp. 376 (in Russian).


\bibitem[17]{Zhiber} N. F. Gareeva  and A. V. Zhiber, {\it  The second order integrals of the hyperbolic
equations and evolutionary equations}, in Proceedings of the International Conference
"Algebraic and analytic methods in the theory of the differential equations",
1996, Orel, edited by A.G.Meshkov, pp.39-42.

\bibitem[18]{ZhS} A. V. Zhiber and V. V. Sokolov, {\it  Exactly integrable hyperbolic equations of Liouville type},
 Russian Mathematical Surveys, 56(1), 61  (2001).

\bibitem[19]  {Lain} M. E. Lain$\acute{e}$, {\it Sur une $\acute{e}$quation de la forme $s = p\phi(x; y; z; q)$ integrable par
la m$\acute{e}$thode
de Darboux}, Comptes rendus, V. 183, 1926, P. 1254-1256.

\bibitem[20]  {Vessiot39} E. Vessiot,  {\it Sur les $\acute{e}$quations aux d$\acute{e}$riv$\acute{e}$s
partielles du second order,
$F(x; y; z; p; q; r; s; t) = 0$, integrable par la m$\acute{e}$thode de Darboux},  J. Math. pure
appl., V. 18, 1939. P. 1-61.


\bibitem[21]  {Kaptsov} O. V. Kaptsov, {\it  On the Goursat classification problem},
Programming and Computer Software 38 (2), 102-104.

\bibitem[22] {ZhZh1} K. Zheltukhin  and  N. Zheltukhina, {\it On existence of an $x$ - integral for a semi-discrete chain of hyperbolic type}, Journal of Physics Conference Series 670(1): 012055, January 2016.

\bibitem[23]  {ZhZh2} K. Zheltukhin and  N. Zheltukhina, {\it Semi-discrete hyperbolic equations admitting five dimensional characteristic $x$-ring}, Journal of Nonlinear Mathematical Physics 23(3), April 2016.

\end{thebibliography}
\end{document}